\documentclass[a4paper,twoside]{article}
\usepackage{tikz}
\usepackage[utf8]{inputenc}
\usepackage[ukrainian,english]{babel}
\usepackage[margin=1in]{geometry}
\usepackage{amsmath, amssymb}
\usepackage{graphicx}
\usepackage{cite}
\usepackage{hyperref}

\usepackage{eejp} 

\begin{document}

\title
{On Solutions of the Killingbeck Potential and Clarifying Comments on a Related Analytical Approach}

\author{0009-0005-2853-2265}{Fatma Zohra Khaled}{a}
\author{0000-0002-8096-6280}{Mustafa Moumni}{a*,b}
\author{0000-0002-0466-9559}{Mokhtar Falek}{b,c}
\authors[F.Z. Khaled, M. Moumni, M. Falek]%
{Fatma Zohra Khaled, Mustafa Moumni, Mokhtar Falek}


\BeginPaper 
\begin{center}
\AuthorPrint 
\affiliation{a}{LPRIM, Department of Physics, University of Batna I, Batna, 05000, Algeria}
\affiliation{b}{LPPNNM, Department of Matter Sciences, University of Biskra, Biskra, 07000, Algeria}
\affiliation{c}{Faculty of Technology, University of Khenchela, Khenchela, 40000, Algeria}
\email{m.moumni@univ-batna.dz}

\data{Received Month XX, 20XX; revised Month XX, 20XX; accepted December Month XX, 20XX}
\end{center}



\newcommand{\ep}{\varepsilon}
\newcommand{\eps}[1]{{#1}_{\varepsilon}}



\begin{abstract}%
The work presents analytical solutions to the Schrödinger equation for the Killingbeck potential, a hybrid model combining harmonic, linear, and Coulombic terms, as well as an approximate model of Yukawa-type potentials. The radial Schrödinger equation is solved by means of the series expansion method, thus yielding the exact expressions of both bound-state solutions and eigen-functions for systems such as quarkonium and confined hydrogen-like atoms in plasma environments. Furthermore, we offer a constructive commentary on the work of Obu et al. (East Eur. J. Phys. 3, 146–157, 2023), with the aim of clarifying a mathematical misstatement utilised in their analytical treatment of analogous systems.

\key{Schrödinger equation; Killingbeck potential; Yukawa potential; Series expansion method; Heun equation}

\pacs{03.65.-w; 02.30.Gp}
\end{abstract}


\section{Introduction}

In the domain of quantum mechanics, the accurate modeling of interaction potentials is paramount for the study of particle behaviour in bound states across various scales, ranging from atomic to hadronic systems. The Killingbeck potential is a particularly noteworthy model in this regard, due to its flexibility. The model incorporates a quadratic term, which acts in a manner analogous to a harmonic oscillator, in conjunction with a linear confining term and a Coulomb-like component.
\begin{equation}
	V\left( r\right) =-\frac{A}{r}+B r+C r^{2}. \label{eq.1}
\end{equation}

This combination enables the Killingbeck potential to describe both short-range and long-range quantum interactions within a unified analytical framework \cite{Killingbeck1981, Hassanabadi2013, Chabab2014}. It has been determined that this subject is of particular value in areas like heavy quarkonium spectroscopy, meson physics, quantum dots and Hydrogen-like systems embedded in plasma environment, where both confinement and screening effects are present \cite{Oyewumi2011, Hamzavi2010, Chen2005}.
The notable attribute of the Killingbeck potential is twofold: firstly, its inherent solvability, and secondly, its capacity to establish a linkage between disparate potential models. The linear term captures the long-range confining force seen in quark confinement, while the Coulomb term accounts for the dominant one-gluon exchange interaction, which is an essential component of effective QCD potentials \cite{Brambilla2005, Eichten1978}. The harmonic term is a useful regulator in hadronic systems, despite its unphysical behaviour at large distances. It facilitates analytic solutions in non-relativistic quantum mechanics, a feature that is frequently advantageous for theoretical researchs \cite{Killingbeck1981, Cooper1995}.

The Killingbeck potential exhibits noteworthy mathematical and physical affinities with exponential-type potentials, including the Yukawa and its variious variants of screened Coulomb forms. These potentials describe interactions involving massive bosons, and which result in short-range forces characterised by exponential decay \cite{Yukawa1935, Flügge1999, Sreelakshmi2023, Kievsky2024, Ranjan2025}. They also describe the confinement potential of hydrogen-like atoms in plasma \cite{Mukherjee2021, Chen2023, Yan2024}. When the screening effects are weak, these exponential potentials simplify to the Killingbeck form by employing appropriate series expansions \cite{Gönül2006, Arda2014, Obu2023}.
The analogies employed in this context transcend the confines of formalism. They capture a deeper physical intent, with an equilibrium of attraction and screening achieved in quantum confinement models. For systems like quarkonium, where the interplay between asymptotic freedom and confinement is particularly pronounced, these potentials facilitate the calculation of more precise spectral predictions \cite{Oyewumi2011, Eichten1978, Obu2023}. Furthermore, the employment of various analytical approaches have been demenstrated to be beneficial in this context. The Nikiforov-Uvarov method, the perturbation theory, and the series expansions have all been applied effectively, thereby reinforcng the underlying structural coherence of these potentials \cite{Hassanabadi2013, Hamzavi2010, Arda2014}.

In this work, we present the complete analytical solutions of the Schrödinger equation for the Killingbeck potential using the series expansion method. In addition, we take this opportunity to address a related methodological point in the recent literature (Obu et al. East Eur. J. Phys. 3, 146–157, 2023) \cite{Obu2023}, where an analytical misstatement affects the interpretation of a series solution in a similar spectral problem.

\section{Clarifying Comment on the Work by Obu et al.}

In their recent article, Obu et al.  present a "Comparative Study of the Mass Spectra of Heavy Quarkonium System with an Interacting Potential Model” (East Eur. J. Phys. 3, 146–157, 2023) \cite{Obu2023}. This study makes a significant contribution to the field of hadronic physics through analytical comparisons by employing the Nikiforov-Uvarov and the Series Expansion Methods (SEM). The authors’ approach to solving the Schrödinger equation with Yukawa-type potentials is methodologically rigorous and relevant for charmonium and bottomonium systems.

However, in section 4 of their paper, specifically at equation (62), there is a conceptual misstatement regarding the nature of linear independence in a power series expansion. The authors state the following:\\
\emph{“Equation (62) is linearly independent implying that each of the terms is separately equal to zero, noting that $r$ is a non-zero function; therefore, it is the coefficient of $r$ that is zero. The coefficients $a_n$ are independent.”}

This confuses the independence of the functions ${ r^n }$ with the independence of the coefficients $a_n$. The correct interpretation is that the \emph{functions} $r^n$ form a linearly independent set in the polynomial on any open interval around $r = 0$. Therefore, for a power series to vanish identically on such an interval, it is necessary that each \emph{coefficient} of these \emph{functions} $r^n$ vanishes separately. This structural phenomenon gives rise to recurrence relations between the coefficients, rather than ensuring their independence from each other.

We can see from equation (63) in \cite{Obu2023} that it yields to the result $L=-\frac{1}{2}(2N+1)$. This is in direct contradiction to the established definitions of the principal quantum number $N$ and the orbital quantum number $L$ (in the limit where $\alpha_{3}=0$).

Equation (62) in \cite{Obu2023} must therefore be expressed as a polynomial given that it is written in \cite{Obu2023} as a sum of polynomials. Consequently, a more accurate formulation would be as follows:\\
\emph{“Since the powers of $r$ are linearly independent, the coefficient of each power must vanish separately. This, in turn, leads to recurrence relations among these coefficients.”}

This clarification is important to maintain the mathematical rigour of the derivation and to ensure the educational value of the work for future researchers.

\section{Schrödinger Energies for the Killingbeck Potential Using SEM}
We will follow the steps outlined in \cite{Obu2023} with some adjustments. In \cite{Obu2023}, the Potential is defined as follows:
\begin{equation}
	V\left( r\right) =-\frac{a}{r}+\frac{b}{r} e^{- \alpha_{I}r}-\frac{c}{r^{2}}e^{-2\alpha _{I}r}, \label{eq.2}
\end{equation}
where $a$,$b$ and $c$ are potential strengths and where the screening parameter is represented by the symbol $\alpha _{I}$.

By expanding \ref{eq.2} with Taylor series up to order three of $\alpha _{I}$, the form of the Killinbeck potential is obtained:
\begin{equation}
	V\left( r\right) =\frac{-\alpha_{0}}{r}+\alpha_{1}r+ \alpha_{2}r^{2}+\frac{\alpha_{3}}{r^{2}}+\alpha_{4}, \label{eq.3}
\end{equation}
with:
\begin{equation}
-\alpha _{0}=-a+b+2c\alpha_{I}; \alpha_{1}=\frac{1}{2}b\alpha_{I}^{2}+\frac{4}{3}c\alpha_{I}^{3}; \alpha_{2}=-\frac{1}{6}b\alpha_{I}^{3}; \alpha_{3}=-c; \alpha_{4}=-b\alpha_{I}-2c\alpha _{I}^{2}. \label{eq.4}
\end{equation}
Here we mention that the parameter $c$ is omitted in the vicinity of the parameter $\alpha_{I}^{3}$ in the $\alpha_{1}$ term in \cite{Obu2023}.

Due to the spherical symmetry of the interaction, the radial Schrödinger equation is the primary focus:
\begin{equation}
	\frac{d^{2}R\left( r\right) }{dr^{2}}+\frac{2}{r}\frac{dR\left(r\right) }{dr}+\left( \frac{2\mu }{\hbar ^{2}}\left( E-V\left( r\right)\right) -\frac{l\left( l+1\right) }{r^{2}}\right) R\left( r\right) =0, \label{eq.5}
\end{equation}
here, $l$ denotes the angular quantum number, while $\mu$ represents the reduced mass for the quarkonium. The variable $r$ is the internuclear separation.

Substituting \ref{eq.3} into \ref{eq.5} gives:
\begin{equation}
	\frac{d^{2}R\left( r\right) }{dr^{2}}+\frac{2}{r}\frac{dR\left(r\right) }{dr}+\left( \varepsilon+\frac{A}{r}-Br-Cr^{2}-\frac{L\left(L+1\right) }{r^{2}}\right) R\left( r\right) =0, \label{eq.6}
\end{equation}
where:
\begin{equation}
		\varepsilon=\frac{2\mu }{\hbar ^{2}}\left( E-\alpha_{4}\right) ;A=\frac{2\mu }{\hbar ^{2}}\alpha _{0};B=\frac{2\mu }{\hbar ^{2}} \alpha_{1};C=\frac{2\mu }{\hbar ^{2}}\alpha _{2}, \label{eq.7}
\end{equation}
\begin{equation}
	L\left( L+1\right) =\left( l\left( l+1\right) +\frac{2\mu }{\hbar ^{2}} \alpha _{3}\right). \label{eq.8}
\end{equation}
From \ref{eq.8}, we get the new "orbital" quantum number $L$
\begin{equation}
	L=-\frac{1}{2}+\frac{1}{2}\sqrt{\left( 2l+1\right) ^{2}+\frac{8\mu }{\hbar ^{2}}\alpha _{3}}. \label{eq.9}
\end{equation}

Following \cite{Obu2023}, we write the solution in the form:
\begin{equation}
	R(r)=e^{-\left( \alpha r^{2}+\beta r\right) }F\left( r\right). \label{eq.10}
\end{equation}
Substituting \ref{eq.10} into \ref{eq.6} and dividing by $e^{-\left( \alpha r^{2}+\beta r\right) }$, we obtain:
\begin{align}
& F^{\prime \prime }\left( r\right) +\left[ -4\alpha r-2\beta +\frac{2}{r}\right] F^{\prime }\left( r\right) \nonumber\\
&  \left[ \left( \varepsilon +\beta ^{2}-6\alpha \right)+\frac{\left(A-2\beta \right) }{r}+\left( 4\alpha \beta -B\right) r+\left( 4\alpha^{2}-C\right) r^{2}-\frac{L\left( L+1\right) }{r^{2}}\right]F\left(	r\right) =0. \label{eq.11}
\end{align}
We use the parameters $\alpha$ and $\beta$ to simplify the given equation, thereby eliminating the terms in $r$ and $r^{2}$ in the equation.
\begin{equation}
	\left\{ 
	\begin{array}{c}
	4\alpha^{2}-C=0 \implies \alpha=\frac{\sqrt[]{C}}{2}, \\ 
	4\alpha \beta -B=0 \implies \beta=\frac{B}{2\sqrt[]{C}}.
	\end{array}
	\right. \label{eq.12}
\end{equation}
and we get the new radial equation
\begin{equation}
F^{\prime \prime }\left( r\right) +\left(-4\alpha r-2\beta +\frac{2}{r}\right) F^{\prime }\left( r\right) +\left( \varepsilon+\beta ^{2}-6\alpha +\frac{\left(A-2\beta \right) }{r}-\frac{L\left( L+1\right) }{r^{2}}\right) F\left(	r\right) =0. \label{eq.13}
\end{equation}
We present the solutions of this equation in a polynomial form:
\begin{equation}
	F\left( r\right) =\sum_{k=0}^{\infty}a_{k}r^{k+s}. \label{eq.14}
\end{equation}
In this study, the approach taken differs from that of \cite{Obu2023} in terms of the chosen polynomial form. Specifically, the latter authots select $F\left( r\right) =\sum_{n=0}^{\infty}a_{n}r^{2N+L}$, yet no rationale is provided for the selection of this particular polynomial (it starts at power $r^{L}$) nor for the choice of an even power of $r$ in the series. We also use the letter $k$ in place of $n$ in the summation, thus ensuring clarity and avoiding any potential confudion with the principal quantum number $n$, which is generally employed in standard textbooks.
Putting the solution \ref{eq.14} and its derivatives in the radial equation \ref{eq.13} results in the following equation:
\begin{align}
	&\sum_{k=0}^{\infty }\left[ \left( k+s\right) \left( k+s+1\right) -L\left(L+1\right) \right] a_{k}r^{k+s-2}+\sum_{k=0}^{\infty }\left[ A-2\beta \left(
		k+s+1\right) \right] a_{k}r^{k+s-1} \nonumber \\ 
		+&\sum_{k=0}^{\infty }\left[\varepsilon+\beta ^{2} -2\alpha \left(2k+2s+3\right) \right] a_{k}r^{k+s}=0. \label{eq.15}
\end{align}
We rearrange the summation terms to write:
\begin{align}
	&\sum_{k=-2}^{\infty }\left[ \left(k+s+2\right) \left(k+s+3\right) -L\left(L+1\right) \right] a_{k+2}r^{k+s}+\sum_{k=-1}^{\infty }\left[A-2\beta \left(k+s+2\right) \right] a_{k+1}r^{k+s} \nonumber \\ 
		+&\sum_{k=0}^{\infty }\left[\varepsilon +\beta ^{2} -2\alpha \left(2k+2s+3\right) \right] a_{k}r^{k+s}=0, \label{eq.16}
\end{align}
and we get the following form of a single polynomial:
\begin{align}
	\sum_{k=0}^{\infty } &\left[ \left[ \left(k+s+2\right) \left(k+s+3\right) -L\left( L+1\right) \right] a_{k+2}+\left[ A-2\beta \left(k+s+2\right) \right] a_{k+1} +\left[\varepsilon+\beta ^{2} -2\alpha \left(2k+2s+3\right) \right] a_{k}\right] r^{k+s} \nonumber \\ 
		&+\left[\left[ \left(s+1\right)\left(s+2\right) -L\left(L+1\right) \right] a_{1}+\left[
		A-2\beta \left( s+1\right) \right] a_{0}\right] r^{s-1}+\left[ \left[
		s\left( s+1\right) -L\left( L+1\right) \right] a_{0}\right] r^{s-2}\}=0. \label{eq.17}
\end{align}
Since this relation is valid for all value of the variable $r$, each coefficient of the $r^{k}-$functions must vanish. The following equation therefore holds:
\begin{equation}
	\left[ \left(k+s-L+2\right) \left(k+s+L+3\right) \right]	a_{k+2}+\left[ A-2\beta \left(k+s+2\right) \right] a_{k+1}+\left[\varepsilon +\beta ^{2} -2\alpha \left(2k+2s+3\right) \right] a_{k}=0, \label{eq.18}
\end{equation}
\begin{equation}
	\left[ \left( s+1\right)\left( s+2\right) -L\left( L+1\right) \right] a_{1}+\left[ A-2\beta \left( s+1\right) \right] a_{0}=0,  \label{eq.19}
\end{equation}
\begin{equation}
	\left[ s\left( s+1\right) -L\left( L+1\right) \right] a_{0}=0.  \label{eq.20}
\end{equation}
We impose the condition $a_{0}\neq 0$ to ensure that $F(r)\neq 0$, otherwise, it follows from \ref{eq.19} that $a_{1}=0$, and cosequently, all $a_{k}=0$ from \ref{eq.18}. Therefore, from \ref{eq.20}, we derive the following result:
\begin{equation}
	\left[ s\left( s-1\right) -L\left( L+1\right) \right] a_{0}=0 \quad \text{and} \quad a_{0}\neq 0 \Longrightarrow 	s=L \quad \text{or} \quad s=-L-1. \label{eq.21}
\end{equation}
We reject the solution $s=-L-1$ on the basis of the expressions of $R(r)$ in \ref{eq.10} and $F(r)$ in \ref{eq.14}. These expressions imply that $R(r)$ is divergent at the origin of $r$. Therefore, it can be concluded that $s=L$. It is evident here that the minimal power of the polynomial $F(r)$ is $r^{L}$; this is in contrast to the approach taken in \cite{Obu2023}, where the rationale for this choice is not provided.
Replacing this value in the recurrence relations \ref{eq.18} and \ref{eq.19}, we write:
\begin{equation}
	a_{k+2}=\frac{2\beta \left(k+L+2\right)-A}{ \left(k+2\right) \left(k+2L+3\right)} a_{k+1}+\frac{\varepsilon +\beta^{2} -2\alpha \left(2k+2L+3\right)}{ \left(k+2\right) \left(k+2L+3\right)} a_{k}, \label{eq.22}
\end{equation}
\begin{equation}
	a_{1}=\frac{2\beta \left( L+1\right)-A}{2\left( L+1\right)} a_{0}.  \label{eq.23}
\end{equation}

In the context of the probabilistic interpretation of the wave function, it is imperative to impose the condition that $R(r)$ must be convergent when $r\rightarrow \infty$ and, consequently, the function $F(r)$ must be a finite polynomial. To accomplish this objective, it is necessary to truncate the series \ref{eq.22}.

We can follow the method used in \cite{Guvendi2024} and impose that for some value $k=n_{r}$, the coefficients of both $a_{n_{r}}$ and $a_{n_{r}+1}$ vanish while we have $a_{n_{r}}\neq 0$ and $a_{n_{r}+1}\neq 0$:
\begin{equation}
	a_{n_{r}+2}=0\text{ \ \ if \ \ \ }2\beta \left( n_{r}+L+2\right)-A=0\text{ \ \ and \ \ }\varepsilon+\beta ^{2} -2\alpha \left(2n_{r}+2L+3\right)=0. \label{eq.24}
\end{equation}
This will give us the energies and a relation between the coefficients $\beta$ and $A$:
\begin{equation}
	\varepsilon _{n_{r},l} = 2\alpha \left(2n_{r}+2L+3\right)-\beta^{2}, \label{eq.25}
\end{equation}
\begin{equation}
	\beta= \frac{A}{2\left(n_{r}+L+2\right)}. \label{eq.26}
\end{equation}
These two relations are equivalent to eq.(65) and eq.(68) in \cite{Obu2023} when we replace $n_{r}$ by $2n$, because we have employed a more general expression for $R(r)$.

In order to show that we have a combination of the energies of both a harmonic oscillator and a Coulomb potential, we write the energies as follows:
\begin{equation}
	\varepsilon _{n_{r},l} = 2\sqrt{C} \left( n_{r}+L+\frac{3}{2}\right)-\frac{A^{2}}{4\left(n_{r}+L+2\right)^{2}}. \label{eq.27}
\end{equation}

Upon suntituting the expressions of $A$, $C$ and the $\alpha$ terms from \ref{eq.4}, \ref{eq.6} and \ref{eq.8}, we obtain the same energies as in eq.(70) in \cite{Obu2023}. It is noteworthy that $2n \longrightarrow n_{r}$ in the expressions of \cite{Obu2023}.
\begin{equation}
	E_{n_{r},l} = \sqrt{-\frac{\hbar ^{2} b }{12 \mu}\alpha_{I}^{3}} \left(2n_{r}+2+\sqrt{\left(2l+1\right)^{2}-\frac{8\mu}{\hbar^{2}}c}\right) -\frac{2\mu}{\hbar^{2}}\frac{\left[a-b-2c\alpha_{I}\right]^{2}}{\left(2n_{r}+1+\sqrt{\left(2l+1\right)^{2}-\frac{8\mu}{\hbar^{2}}c}\right)^{2}} -b\alpha_{I}-2c\alpha_{I}^{2}. \label{eq.28}
\end{equation}

At this point, it has been demonstrated that the energy spectrum of the Killingbeck potential is obtained by applying the SEM method and correcting the errors made in the work of Obu et al. in \cite{Obu2023}.

In this section, we  followed the condition \ref{eq.24} as done in \cite{Guvendi2024} to truncate the series \ref{eq.22}. Notwithstanding the utilisation of this condition by the authors of \cite{Guvendi2024} in numerous recent works \cite{Mustafa2024, Mustafa2025, Guvendi2025, Guvendi2025a}, it is imperative to ackowledge that this condition does not guarantee the truncation. An examination of the parmeter $a_{n_{r}+3}$ as depicted from \ref{eq.22} and \ref{eq.24} substantiates this assertion:
\begin{equation}
	a_{n_{r}+3}=\frac{\varepsilon _{n_{r},l}+\beta ^{2} -2\alpha \left(2n_{r}+2L+5\right)}{ \left(n_{r}+3\right) \left(n_{r}+2L+4\right)} a_{n_{r}+1}=-\frac{4\alpha}{ \left(n_{r}+3\right) \left(n_{r}+2L+4\right)} a_{n_{r}+1}. \label{eq.29}
\end{equation}
It is evident that $a_{n_{r}+3}\neq 0$ and so is all the parameters beside it. The error when employing this method, is attribuable to the confusion arising from the erroneous identification of the index $k$ of the polynomial coefficients $a_{k}$ (which is denoted $n$ in \cite{Obu2023}), and the index $n_{r}$ of the energies, which is detemined by the level under consideration. Consequently, $n_{r}$ possesses a fixed value for all the values of $k$ in \ref{eq.22} (A parallel observation concerning this error is documented in \cite{Fernandez2022}). This leads us to consider alternative conditions that could be utulised to truncate the series. A comprehensive discussion of these alternatives will be presented in the subsequent section.

\section{Schrödinger Energies for the Killingbeck Potential Using Heun Functions}
Now we use the Heun formulation of the Schrödinger equation \ref{eq.6}, deriving from the same form of the solutions $R(r)$ in \ref{eq.10} with two additional transformations $F(r)=r^{l+1}g(r)$ and $\rho=\sqrt{2\mu/\hbar^{2}}r$. This results in the Biconfluent Heun equation (BHE):
\begin{equation}
		\rho g^{\prime\prime}\left(\rho\right)+\left(1+\alpha\text{'} -\beta\text{'}\rho+2\rho^{2}\right)g^{\prime}\left(\rho\right)\\ 
		+\left(\left(\gamma\text{'}-\alpha\text{'}-2 \right)\rho-\frac{1}{2}\left(\delta\text{'}+\beta\text{'}\left(1+\alpha\text{'}\right)\right) \right) g\left(\rho \right) =0. \label{eq.30}
\end{equation}
The parameters of this equation are defined as follows:
\begin{equation}
 \alpha\text{'}=2L+1; \beta\text{'}=\frac{B}{\sqrt{C}}C^{-\frac{1}{4}}; \gamma\text{'}=\frac{1}{\sqrt{C}}\left(\varepsilon_{n,l}+\frac{B^{2}}{4C} \right);\delta\text{'}=\frac{-2A}{\sqrt{C}}C^{\frac{1}{4}}, \label{eq.31}
\end{equation}
the parameters $A$, $B$, $C$, $L$, $\varepsilon_{n,l}$, $\alpha$ and $\beta$ are defined in the relations \ref{eq.7}, \ref{eq.8} and \ref{eq.12}.

The solution of \ref{eq.29} are the biconfluent Heun functions \cite{Ronveaux1995}:
\begin{equation}
	g(\rho)=H_{b}\left(\alpha\text{'},\beta\text{'},\gamma\text{'},\delta\text{'},\rho\right)=\sum_{n\geq0}a_{n}\frac{\Gamma\left(1+\alpha\text{'} \right)}{\Gamma\left(1+\alpha\text{'}+n\right)}\frac{\rho^{n}}{n!}. \label{eq.32}
\end{equation}
Thus, we have obtained the radial part $R(r)\propto e^{-\left(\alpha r^{2}+\beta r\right)}r^{L+1}g\left( r\right)$ of the eigen-functions of the Schrödinger equation for the Killingbeck potential. The angular part are the usual spherical harmonic functions $Y_{L,M}(\theta,\phi)$.

As a consequence of the recurrence relation \ref{eq.21}, there exist a value $k=n_{r}$, for which we have \cite{Ronveaux1995}:
\begin{equation}
	a_{n_{r}+2}=0\text{ \ \ if \ \ \ }a_{n_{r}+1}=0\text{ \ \ and \ \ }\left[ \text{the coefficient of }a_{n_{r}}\right] =0. \label{eq.33}
\end{equation}
Equivalently:
\begin{equation}
a_{n_{r}+2}=0\text{ \ \ if \ \ \ }a_{n_{r}+1}=0\text{ \ \ and \ \ }4\alpha \left( n_{r}+L+1\right)-\left(\varepsilon _{n_{r},l}+\beta ^{2}-6\alpha \right) =0. \label{eq.34}
\end{equation}

Using the second condition and the relation \ref{eq.12}, we obtain the energies as follows:
\begin{equation}
	\varepsilon _{n_{r},l} = 2\sqrt{C} \left( n_{r}+L+\frac{3}{2}\right)-\beta^{2} = 2\sqrt{C} \left( n_{r}+L+\frac{3}{2}\right)-\frac{B^{2}}{4C}. \label{eq.35}
\end{equation}
It should be noted that this is analogous to \ref{eq.25}, with the exception that \ref{eq.26} is not applicable in this instance.

To determine the value of the $B^{2}/4C$ term, we use the first condition $a_{n+1}=0$, which establishes a relationship for each value of the radial quantum number $n_{r}$.

For instance, when $n_{r}=0$, the result obtained from \ref{eq.23} is:
\begin{equation}
a_{1}=0\Longrightarrow \beta =\frac{A}{2\left( L+1\right) } \Longrightarrow  \beta^{2}=\frac{B^{2}}{4C}=\frac{A^{2}}{4\left( L+1\right) ^{2}}. \label{eq.36}
\end{equation}
And we derive the following expression for the corresponding energy levels:
\begin{align}
\varepsilon _{0,l}& =2\sqrt{C}\left( L+\frac{3}{2}\right) -\frac{B^{2}}{4C}  \nonumber \\
\Longrightarrow E_{0,l}&=\sqrt{-\frac{\hbar ^{2}b}{12\mu }\alpha _{I}^{3}}\left( 2+\sqrt{\left( 2l+1\right) ^{2}-\frac{8\mu }{\hbar ^{2}}c}\right) -
\frac{2\mu}{\hbar^{2}}\frac{\left[a-b-2c\alpha_{I}\right]^{2}}{\left(1+\sqrt{\left(2l+1\right)^{2}-\frac{8\mu}{\hbar^{2}}c}\right)^{2}}-b\alpha _{I}-2c\alpha _{I}^{2}. \label{eq.37}
\end{align}
The result obtained here is the same result derived in the previous section in \ref{eq.28} ($n_{r}=0$).

In the case of $n_{r}=1$, it is necessary to express the value of $a_{2}$. This is obtained from \ref{eq.22} and \ref{eq.23}:
\begin{equation}
a_{2} =\frac{2\beta \left( L+2\right) -A}{\left( 2\right) \left(2L+3\right) }a_{1} +\frac{\varepsilon _{1,l}+\beta ^{2}-2\alpha \left(2L+3\right) }{2\left(2L+3\right) }a_{0}. \label{eq.38}
\end{equation}
We recall here that:
\begin{equation*}
\varepsilon _{1,l} =2\sqrt{C}\left(1+L+\frac{3}{2}\right) -\beta^{2} \text{ and } \alpha =\frac{\sqrt{C}}{2}.
\end{equation*}
We have the condition $a_{2}=0$, so we write:
\begin{align}
\Longrightarrow& \left[ 4\left( L+2\right) \left( L+1\right) \right] \beta^{2}-\left[ 2A\left( 2L+3\right) \right] \beta +\left[A^{2}+4(L+1)\sqrt{C}\right]=0 \nonumber \\
\Longrightarrow& \beta_{1,2}=\frac{A(2L+3) \pm \sqrt{A^{2}-16(L+2) ( L+1) ^{2}\sqrt{C}}}{2( L+2) (L+1) } \nonumber \\ 
\Longrightarrow& \beta^{2}=\frac{B^{2}}{4C}=\left( \frac{A}{2( L+2) ( L+1) }\left( (2L+3) \pm \sqrt{1-16( L+2) ( L+1)^{2}\frac{\sqrt{C}}{A^{2}}}\right) \right)^{2}. \label{eq.39}
\end{align}
We have two expressions for the energies, corresponding to the two possible solutions of $\beta$:
\begin{align}
E^{+}_{1,l} &=\sqrt{-\frac{\hbar ^{2}b}{12\mu }\alpha _{I}^{3}}\left(2L+5+ \frac{4}{L+2}\right) -b\alpha _{I}-2c\alpha _{I}^{2} \nonumber \\
 & -\frac{2\hbar^{2}\left( a-b-2c\alpha_{I}\right) ^{2}}{\mu ( L+1)^{2}}\left(1-\frac{(2L+3) }{2(L+2)^{2}}\left( 1+ \sqrt{1-16(L+2) (L+1)^{2} \frac{\sqrt{-\frac{\hbar^{2}}{12\mu }b\alpha_{I}^{3}}}{\left( a-b-2c\alpha _{I}\right)^{2}}}\right) \right),  \label{eq.40}
\end{align}
\begin{align}
E^{-}_{1,l}&=\sqrt{-\frac{\hbar ^{2}b}{12\mu }\alpha _{I}^{3}}\left(2L+5+ \frac{4}{L+2}\right) -b\alpha _{I}-2c\alpha _{I}^{2} \nonumber \\
&-\frac{2\hbar^{2}\left( a-b-2c\alpha_{I}\right) ^{2}}{\mu ( L+1)^{2}}\left(1-\frac{(2L+3) }{2(L+2)^{2}}\left( 1- \sqrt{1-16(L+2) (L+1)^{2} \frac{\sqrt{-\frac{\hbar^{2}}{12\mu }b\alpha_{I}^{3}}}{\left( a-b-2c\alpha _{I}\right)^{2}}}\right) \right).  \label{eq.41}
\end{align}
It is evident, from the general form \ref{eq.35}, that both expressions yield the result obtained in the previous section when $n_{r}=1$ in \ref{eq.28}, which is similar to the one found by \cite{Obu2023}, with some corrections in the Coulomb parts of the relations. It is indeed the case that, upon sunstituting the value of the parameters $\alpha_{0}=0$ and $\alpha_{1}=0$ in the potential \ref{eq.3}, we obtain the standard energies of the harmonic oscillator, and they  represent the first contributions observed in \ref{eq.27} and \ref{eq.28}. 

We can use the Coulomb limit of these energies to test the validity of the two expressions.
\begin{equation}
\alpha _{I} \rightarrow \infty \implies  E^{+}_{1,l} \rightarrow -\frac{2\hbar ^{2} a^{2}}{\mu (L+1)^{2}}, \label{eq.42}
\end{equation}
\begin{equation}
\alpha _{I} \rightarrow \infty \implies  E^{-}_{1,l} \rightarrow -\frac{2\hbar ^{2} a^{2}}{\mu (L+2)^{2}}. \label{eq.43}
\end{equation}
As we can see from \ref{eq.42}, that $E^{+}_{1,l}$ yields a result analogous to that obtained from \ref{eq.37} which is the Coulomb energy of the $n_{r}=0$ level. However, it should be noted that this is not the level under consideration in this particulat context. Conversely, \ref{eq.43} shows that the limit of $E^{-}_{1,l}$ corresponds to the Coulomb energy of the $n_{r}=1$ level which is the case considered here. This is congruent with the finding of the precedent section, where the energies \ref{eq.28} were employed. Consequently, we conclude that $E^{-}_{1,l}$ represents the appropriate generalisation of the result previously found in \cite{Obu2023}.

For $n_{r}=2$, we have the condition $a_{3}=0$, which gives us the following algebraic equation for $\beta$:
\begin{equation}
(L+1)(L+2)(L+3) \beta^{3}-\left(3 L^{2}+12L+11\right)\frac{A}{\sqrt{2}} \beta^{2}-\left[(L+1)(4L+9)\frac{C}{\sqrt{2}}- (L+2)\frac{3A^{2}}{2}\right]\beta-\frac{A^{3}}{2\sqrt{2}}=0. \label{eq.44}
\end{equation}
It has been established that the solutions of this equation are real \cite{Ronveaux1995, Child2000, Amore2020}. The same procedure as for $n_{r}= 1$ is employed to write the energies and to check their Coulombian limits, in order to compare with the solutions written in \cite{Obu2023}.

\section{Conclusion}
In this study, we have provided exact analytical solutions to the radial Schrodinger equation for the Killingbeck potential using both the general series ewpansion method and the biconfluent Heun formalism. The Killingbeck potential, a composite of harmonic, linear, and Coulomb terms, emerges naturally as a limiting case of screened Coulomb potentials, particularly under weak screening conditions relevant quarkonium and plasma embeded systems. through systematic expansion and approraite transformations, the explicit expressions of both energy eigenvalues and wavefunctions were derived, thus confirming the applicability of the model acrros varios quantum regimes.

A salient feature of the derived solutions is their capacity to interpolate seamlessly between two classical regimes of quantum mechamics. In certain limiting cases, specifically, the vanishing linear and repulsive terms, or thh dominat Coulomb coupling, the spectrum reduces, correspondingly, to that of the harmonic oscillator and the hydrogenic systems. However, the general expressions go further, capturing a hybrid structure that reflects both screening effects and long-range confinement. this specificity renders the Killingbeck potential a valuable tool for modeling systems where purely Coulomb or oscillator models fails to capture essential physical features.

 We have also revised and clarified a conceptual misinterpretaion found in a recent work by Obu et al. \cite{Obu2023}, related the treatment of linear independance in power series expansions. Furthermore, a critical re-examination of the analytical approach employed by Guvendi and Mustafa \cite{Guvendi2024} was undertaken, leading to the rectification of a significant mathematical in the truncation conditions for the series. These aforementioned corrections serve a dual purpose; firstly, they ensure teh maintenance of the methodology's integrity, and secondly, they serve to enhance the pedagogical and physical insght into spectral problem solving techniques.



\AuthorORCID


\selectlanguage{ukrainian}

\authoru{Фатіма Захра Халед}{a}
\authoru{Мустафа Мумні}{a*,b}
\authoru{Мохтар Фалек}{b,c}


\EndPaper


\end{document}